\documentstyle[color,aps,prl,psfig,epsfig]{revtex}
\begin{document}
\def\bdm{\begin{displaymath}}
\def\edm{\end{displaymath}}
\def\nn{\nonumber}
\def\bc{\begin{center}}
\def\ec{\end{center}}
\def\be{\begin{equation}}
\def\ee{\end{equation}}
\def\tcb{\textcolor{blue}}
\def\tcbl{\textcolor{black}}
\def\tcg{\textcolor{green}}
\def\tcr{\textcolor{red}}
\def\tcgr{\textcolor{grey}}
\def\va{{\bf a}}
\def\vA{{\bf A}}
\def\vb{{\bf b}}
\def\vB{{\bf B}}
\def\vb{{\bf b}}
\def\vc{{\bf c}}
\def\vC{{\bf C}}
\def\vd{{\bf d}}
\def\hvd{\hat\vd}
\def\vD{{\bf D}}
\def\ve{{\bf e}}
\def\hve{\hat\ve}
\def\vE{{\bf E}}
\def\vf{{\bf f}}
\def\vF{{\bf F}}
\def\vg{{\bf g}}
\def\vG{{\bf G}}
\def\vh{{\bf h}}
\def\vH{{\bf H}}
\def\vi{{\bf i}}
\def\vI{{\bf I}}
\def\vj{{\bf j}}
\def\vJ{{\bf J}}
\def\vk{{\bf k}}
\def\hvk{\hat\vk}
\def\vK{{\bf K}}
\def\vl{{\bf l}}
\def\vL{{\bf L}}
\def\vLambda{{\bf\Lambda}}
\def\vm{{\bf m}}
\def\vM{{\bf M}}
\def\vn{{\bf n}}
\def\hvn{\hat\vn}
\def\vN{{\bf N}}
\def\vone{{\bf 1}}
\def\vp{{\bf p}}
\def\hvp{\hat\vp}
\def\vP{{\bf P}}
\def\vq{{\bf q}}
\def\vQ{{\bf Q}}
\def\vr{{\bf r}}
\def\vR{{\bf R}}
\def\vs{{\bf s}}
\def\vS{{\bf S}}
\def\vt{{\bf t}}
\def\vT{{\bf T}}
\def\vu{{\bf u}}
\def\vU{{\bf U}}
\def\vv{{\bf v}}
\def\vV{{\bf V}}
\def\vw{{\bf w}}
\def\vW{{\bf W}}
\def\vx{{\bf x}}
\def\vX{{\bf X}}
\def\vy{{\bf y}}
\def\vY{{\bf Y}}
\def\vz{{\bf z}}
\def\v0{{\bf 0}}
\def\hvz{\hat\vz}
\def\vZ{{\bf Z}}
\def\vtau{{\bf \tau}}
\def\e{{\rm e}}
\def\kB{k_{\rm B}}
\def\kF{k_{\rm F}}
\def\EF{E_{\rm F}}
\def\NF{N_{\rm F}}
\def\pF{p_{\rm F}}
\def\Tc{T_{\rm c}}
\def\vvF{v_{\rm F}}
\def\vna{{\bf\nabla}}
\def\vPi{{\bf\Pi}}
%
%
\twocolumn[\hsize\textwidth\columnwidth\hsize
\csname@twocolumnfalse\endcsname
\title{Fermi-liquid based theory for the in-plane magnetic anisotropy \\
in untwinned high-T$_c$ superconductors}
\author{I. Eremin$^1$ and D. Manske$^2$}
\address{$^1$Institut f\"ur Theoretische Physik, Freie Universit\"at
Berlin, Arnimallee 14, D--14195 Berlin, Germany}
\address{$^2$ Max--Planck--Institut f\"ur Festk\"orperforschung,
Heisenbergstrasse 1, D--70569 Stuttgart, Germany}
\date{\today}
\maketitle
\begin{abstract}
Using a generalized RPA-type theory we calculate the in-plane
anisotropy of the magnetic excitations in hole-doped high-$T_c$
superconductors. Extending our earlier Fermi-liquid based 
studies on the resonance
peak by inclusion of orthorhombicity we still find two-dimensional
spin excitations, however, being strongly anisotropic. This reflects
the underlying anisotropy of the hopping matrix elements and of
the resultant superconducting gap function. We compare 
our calculations with new experimental data on {\it fully untwinned}
$\mbox{YBa}_2\mbox{Cu}_3\mbox{O}_{6.85}$ and find good agreement.
Our results are in contrast to earlier interpretations on the
in-plane anisotropy in terms of stripes (H. Mook {\it et al.},
Nature {\bf 404}, 729 (2000)), but reveal a conventional solution to this
important problem.
\end{abstract}
\pacs{74.20.Mn, 74.25.-q, 74.25.Ha}]

Since the discovery of high-$\Tc$ superconductors, its mechanism
is still under debate. Perhaps one of the most important question concerns 
the role played by spin excitations in these materials. 
For example, one scenario of superconductivity in layered cuprates 
suggests that Cooper-pairing is due to an exchange of antiferromagnetic 
spin fluctuations\cite{moria}. In this respect, an understanding of the 
so-called resonance peak observed by inelastic neutron scattering (INS) 
experiments\cite{keimer,bourges} at the antiferromagnetic wave 
vector {\bf Q}$_{AF}$ and energy 
$\omega \approx \omega_{res}$ plays an important 
role in the phenomenology 
of high-T$_c$ superconductors. Among various explanations over last years 
there are
two most probable scenarios for the formation of the resonance peak. 
The first one suggests that the two-dimensional CuO$_2$ layers are 
intrinsically unstable towards a stripe formation 
with one-dimensional spin and charge order\cite{emery}. In this picture,
the resonance excitations can be interpreted in terms of 
excitation spectra in a bond-centered stripe state with 
long-range magnetic order\cite{vojta,seibold}. More technically, the   
resonance peak in the stripe-ordered phase corresponds to a saddle point 
in the dispersion of the magnetic excitations.
In the other approach that is a conventional Fermi-liquid one, 
the resonance peak arises as a particle-hole excitation (or spin density wave 
collective mode) in a $d_{x^2-y^2}$-wave superconductor and is a result 
of the strong feedback of the superconductivity on 
the dynamical spin susceptibility below T$_c$ 
\cite{dmiekhb01,onufrieva,chubuk1,morr,prelovsek,li}. 

In order to distinguish between both pictures, a detailed analysis of 
untwinned cuprates is necessary. Recently, INS study in the fully 
untwinned high-temperature superconductor YBa$_2$Cu$_3$O$_{6.85}$ reveals 
two-dimensional character of the magnetic fluctuations\cite{hinkov} 
in contrast to the previous conclusions from measurements 
in the partially untwinned samples 
\cite{mook}. 
Here, motivated by recent experiments we analyze the in-plane
anisotropy of the magnetic excitations in hole-doped high-$T_c$
superconductors within a conventional Fermi-liquid and 
generalized RPA-like approach. 
Extending our earlier studies on the resonance
peak by the inclusion of a small orthorhombicity we still find two-dimensional
spin excitations, however being strongly anisotropic, reflecting
the underlying anisotropy of the hopping matrix elements and of
the resultant superconducting gap function.

In order to describe the phenomenology of the superconducting 
cuprates we employ an effective one-band Hubbard Hamiltonian for the 
CuO$_2$-plane
\begin{equation}
H = \sum_{\langle ij \rangle \sigma} t_{ij}
c_{i\sigma}^{\dagger} c_{j\sigma}  + U \sum_{i} n_{i\uparrow} n_{i\downarrow}\quad
, \label{hubbard}
\end{equation}
where $c_{i\sigma}^{\dagger}$ is a creation operator 
of an electron with spin $\sigma$ on site $i$, 
$U$ denotes the on-site Coulomb repulsion,
and $t_{ij}$ is a hopping matrix element in the 
CuO$_2$-plane. Here, we use the six parameter 
fit of the energy dispersion suggested in Ref.\cite{norman} with the 
following chemical potential and hopping amplitudes ($\mu, t_1, ..., t_5$) 
(the units are in eV): 
(0.1197, -0.5881, 0.1461, 0.0095, -0.1298, 0.0069). The lattice constants are 
set to unity.  
To describe the orthorhombic distortions we introduce a  
parameter $\delta_0$ which
leads to an anisotropy in the hopping integrals along and perpendicular  
to the chains in YBa$_2$Cu$_3$O$_{6.85}$(YBCO).
This one-band approach seems
to be justified at least in the optimally-doped cuprates 
because upon hole doping into the
$\mbox{CuO}_2$-plane antiferromagnetism disappears due to
Zhang-Rice singlet formation and quenching of Cu spins. Further
doping then increases the carrier mobility and a system of
strongly correlated quasiparticles occurs\cite{manskefoot1}. 

\begin{figure}[t]
\centerline{\psfig{file=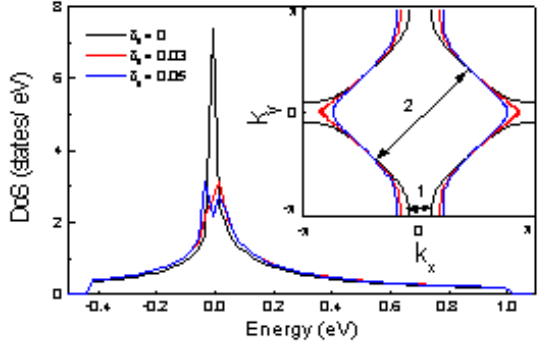,width=7.5cm}}
\caption{(color online) Calculated density of states 
with and without orthorhombicity. For a comparison the corresponding 
changes of the Fermi surface topology are shown in the inset. The arrows refer 
to the two 2{\bf k}$_F$ instabilities as described in the text.}
\label{fig1}
\end{figure}

{\it Normal state}: 
Before analyzing the superconducting state it is instructive 
to understand how the normal state properties and the 
electronic structure of a CuO$_2$ plane 
are affected by the presence of the orthorhombic distortions. 
In Fig.\ref{fig1} we show the calculated density of states 
(DoS) and Fermi surface topology as a function of the orthorhombic distortions.
Without orthorhombicity the DoS reveals a pronounced van-Hove 
singularity (VHS) 
being approximately 19meV below the Fermi level in good agreement with early 
ARPES experiments on YBCO\cite{gofron}. 
Due to the vicinity of the VHS to the Fermi level, 
the effect of the orthorhombic distortions is quite strong. First, at 
$\delta_0=0.03$ the singularity is suppressed and shifted 
slightly above the 
Fermi level. With further increase of $\delta_0$ it splits into 
two peaks. Similar changes occur for the Fermi surface. 
As a function of the orthorhombicity its topology changes and the Fermi 
surface closes around the $(-\pi,0)$ and $(\pi,0)$ points which also leaves an 
impression that the system turns towards a quasi-one-dimensional ones.   
One of the immediate consequence of these changes is that the VHS will be 
present below the Fermi level only around $(0, \pm \pi )$ points. 
This is also consistent with more recent ARPES data on untwinned 
YBCO\cite{schabel} where a suppression of the 
ARPES intensity was observed around $(\pm \pi,0)$  
due to absence of the VHS. However, in Ref.\onlinecite{schabel} 
the change in the Fermi surface topology around $( \pm \pi, 0 )$ was not 
confirmed. It is important to note that this Fermi surface
deformation breaks the point-group symmetry and looks similar to
what is expected for the case of a $d_{x^2-y^2}$-wave Pomeranchuk
instability due to strong electron-electron
interactions \cite{metzner}.

\begin{figure}[t]
\centerline{\psfig{file=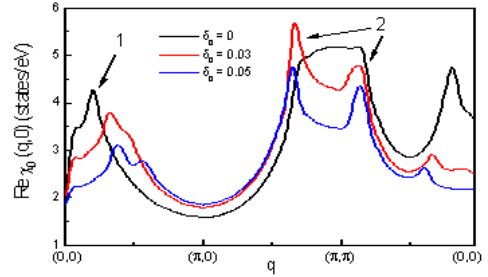,width=6.5cm}}
\caption{(color online) Calculated Re $\chi_0({\bf q},0)$ along the 
path $(0,0) \to (\pi,0) \to (\pi,\pi) \to (0,0)$ for 
various values of the orthorhombic distortions. The arrows indicate the peaks
in Re$\chi_0({\bf q},0)$ arising from 2${\bf k}_{F}$ instabilities as 
shown in Fig.\protect\ref{fig1}.} 
\label{fig2}
\end{figure}

What happens to the spin response if the electronic properties are changed 
due to orthorhombicity? 
We calculate the real part of the bare spin susceptibility in the normal 
state,
\begin{equation}
\chi_0({\bf q},\omega)= \sum_{\bf k} 
\frac{f(\varepsilon_{\bf k}) -f(\varepsilon_{\bf k+q})}
{\varepsilon_{\bf k+q} -\varepsilon_{\bf k} + \omega +i0^+} \quad,
\label{rehi}
\end{equation}
where $f(\varepsilon_{\bf k})$ is the Fermi function. In Fig.\ref{fig2} we 
show the calculated Re$\chi_0 ({\bf q},\omega=0)$ 
as a function of the transferred 
momentum ${\bf q}$. Without orthorhombicity its peak structure reflects two
2${\bf k}_F$ instabilities of the Fermi surface (see inset of Fig.\ref{fig1}). 
The first peak one corresponds to the 
quasi-one-dimensional wave vector connecting the Fermi surface (FS) around 
$(0, \pm \pi)$ points and the other one refers to the 
wave vector connecting the FS along the diagonal of 
the first Brillouin Zone. 
Note, due to strong nesting of the Fermi surface the second structure 
has the form of a plateau. 
If the orthorhombicity is present the first peak is suppressed and 
shifted towards higher ${\bf q}$ values. 
This is due to the fact that the Fermi surface closes around 
$(\pm \pi,0)$ point removing this instability there and moving them 
to higher ${\bf q}$ values around $(0, \pm \pi)$ points. 
On the other hand the diagonal second 2${\bf k}_F$ instability remains mainly 
unchanged and even became more pronounced, since the plateau around 
${\bf Q}_{AF} = (\pi,\pi)$ is suppressed due to the changes of the 
Fermi surface topology at the parts connected by 
this wave vector.

We safely conclude that an orthorhombic distortion change strongly 
the electronic structure of the CuO$_2$-plane and yield characteristic  
changes of the Fermi surface topology, quasiparticle density of states, and 
the spin susceptibility. 
We would like to stress that despite the changes of the Fermi 
surface topology indicating the tendency towards quasi-one-dimensionality, 
the static spin susceptibility remains mainly two-dimensional.

{\it Superconducting state}:
In our one-band model, we further assume that the same quasiparticles 
are participating in the formation of antiferromagnetic
fluctuations and in Cooper-pairing due to these fluctuations. This
leads to the generalized Eliashberg equations which have been
derived and discussed in Refs.
\onlinecite{dmbook,dmiekhb03}. Although these equations
cannot describe the metal-insulator transition properly, we would
like to stress that they allow us to calculate all properties of
the system {\it self-consistently} such as the elementary excitations,
the superconducting order parameter, and the dynamical spin
susceptibility, for example. 

In the pure tetragonal case the resulting superconducting order parameter 
has $d_{x^2-y^2}$-wave symmetry. However, 
in presence of orthorhombicity the superconducting
order parameter changes, since $s$-wave and $d$-wave symmetries 
belong now to the same irreducible representation of the point 
group symmetry. The total superconducting gap 
has the form (weak-coupling limit, $Z=1$)
\begin{equation}
\Delta({\bf k}) = g (\delta_0) \Delta_s +
f(\delta_0)\Delta_d({\bf k})\quad ,\label{gap}
\label{eq:gap}
\end{equation}
\begin{figure}[b]
\centerline{\psfig{file=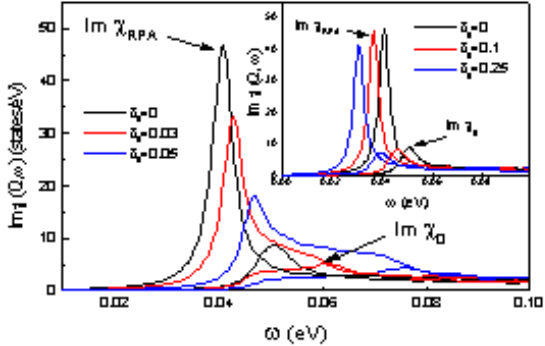,width=7.5cm}}
\caption{(color online) 
Calculated influence of the orthorhombic distortions 
($s$-wave component of the superconducting gap and changes 
in the electronic structure) on the resonance peak. 
Inset: Influence of the admixture of  
$s$-wave superconducting gap {\it only} on the resonance peak formation in the 
tetragonal system. Here, in order to fit the position of 
the resonance at {\bf Q}$_{AF}$ around 41meV we use $U$=0.155eV. 
Note, we further employ the damping $\Gamma=2.4$meV.} 
\label{peak}
\end{figure}
where, for simplicity, we employ $g(\delta_0)=\delta_0$, 
$f(\delta_0)=1-\delta_0$ and $\Delta_d = \Delta_0
(\cos k_x - \cos k_y)/2$, $\Delta_s = \Delta_0$. 
For the set of parameters described
above, we use $\Delta_0 = 26$meV. Note that the additional
$s$-wave component leads to an in-plane anisotropy of the gap
function and thus to different maximum gap values between $(\pm \pi,0)$
and $(0,\pm \pi)$ as observed in Ref. \cite{shen}. 

The most interesting question is: what happens to the dynamical spin 
susceptibility in the superconducting state if an orthorhombic distortion 
is present. Below T$_c$, 
within generalized RPA, the imaginary part of the
dynamical spin susceptibility is given by
\begin{equation}
\mbox{Im}\chi ({\bf q},\omega) = \frac{\mbox{Im}\chi_{0} 
({\bf q},\omega)}{(1-U\mbox{Re}\chi_{0} ({\bf q},\omega))^2 + U^2 
\mbox{Im}\chi^2_{0} ({\bf q},\omega)},
\end{equation}
where $\chi_{0}$ is the BCS Lindhard response function\cite{norman}.
Without orthorhombicity, 
the $d_{x^2-y^2}$ superconducting gap opens rapidly due to a 
feedback
effect on the elementary excitations \cite{letter01} yielding a jump at 
$2\Delta_0$ in Im$\chi_0$ and the resonance condition
\cite{dmiekhb01,onufrieva,chubuk1}
\begin{equation}
1 - U \mbox{Re }\chi_0({\bf q}={\bf Q}_{AF},\omega = \omega_{res}) = 0 
\label{ucr}
\end{equation}
is fulfilled.
Since Im$\chi_0$ is  
zero below $2\Delta_0$, the resonance condition (\ref{ucr})
reveals a strong delta-like peak in Im$\chi$ which occurs only below 
T$_c$. Note, its position is mainly determined by the maximum of the $d$-wave 
superconducting gap $\Delta_0$ and also by the proximity to an  
antiferromagnetic instability described 
by the characteristic energy scale $\omega_{sf}$ (roughly the peak in 
Im $\chi({\bf Q}_{AF},\omega)$ in the normal state). Then, the resonance peak 
scales with the maximum of the $d$-wave 
superconducting gap in 
optimally doped and overdoped compounds. On the other hand, 
in the underdoped cuprates it rather scales with 
$\omega_{sf}$\cite{dmiekhb01,onufrieva,chubuk1} due to stronger 
antiferromagnetic fluctuations. Thus, one finds for the whole doping range  
$\omega_{res}/k_B T_c \approx const$
\cite{dmiekhb01,onufrieva,chubuk1} in good agreement with experiments
\cite{sciencebourges}.

In Fig.\ref{peak} we analyze the influence of the orthorhombic distortions 
on the resonance peak. One clearly sees that the orthorhombicity slightly 
shifts the resonance peak towards higher energies and reduces its intensity 
for increasing $\delta_0$. 
As already mentioned, the position of the resonance peak is determined by the 
strength of the antiferromagnetic fluctuations present in the normal 
state and by the $d$-wave superconducting gap. 
Both are decreasing due to orthorhombicity. 
Namely, as one sees from Fig.\ref{fig2}, the susceptibility 
is decreasing around ${\bf Q}_{AF}$ due to a change of the Fermi 
surface topology. Thus, in the superconducting state 
the resonance is shifted towards higher energies. On the other hand, 
the maximum of the $d-$wave gap is decreasing 
(as we see from Eq.(\ref{eq:gap})) 
which would shift the resonance energy towards lower values. Most importantly, 
we see that the deformation of the electronic structure and Fermi 
surface topology due to orthorhombicity dominate the effect from considering 
solely than the increase of the $s$-wave component 
of the superconducting gap (see inset).  
Noticeable changes occur only for very 
large values of $\delta_0$ when the electronic structure is already strongly 
distorted. Thus we conclude, the observed 
slight shift of the resonance peak position 
in untwinned \cite{hinkov} occurs due to the strong changes in 
the electronic structure induced by the orthorhombic distortion rather than 
due to additional $s$-wave component of the superconducting gap.
\begin{figure}[t]
\centerline{\psfig{file=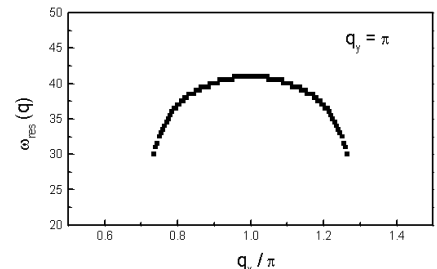,width=5.7cm,angle=0}}
\caption{Calculated dispersion of the
resonance peak towards lower energies for fixed $q_y=\pi$ without including 
the effect of the orthorhombicity.} 
\label{dispersion}
\end{figure}

{\it Comparison with experiment}: Let 
us now turn to the broader analysis of the dispersion of the 
resonance excitations below $\omega_{res}$. 
In the tetragonal case, going away from the points of the Fermi surface 
connected by the antiferromagnetic wave vector ${\bf Q}_{AF}$, we are 
moving towards the diagonal of the BZ where the $d_{x^2-y^2}$-wave 
superconducting gap is zero. As a result the resonance energy, $\omega_{res}$  
shifts towards smaller values. 
This results in the parabolic shape of resonance energy dispersion as 
shown in Fig.\ref{dispersion} calculated for fixed $q_y = \pi$. 
This parabolic behavior obtained in
our calculations agrees well with the experimental findings of Bourges
{\it et al.} \cite{sciencebourges}.

\begin{figure}[t]
\centerline{\psfig{file=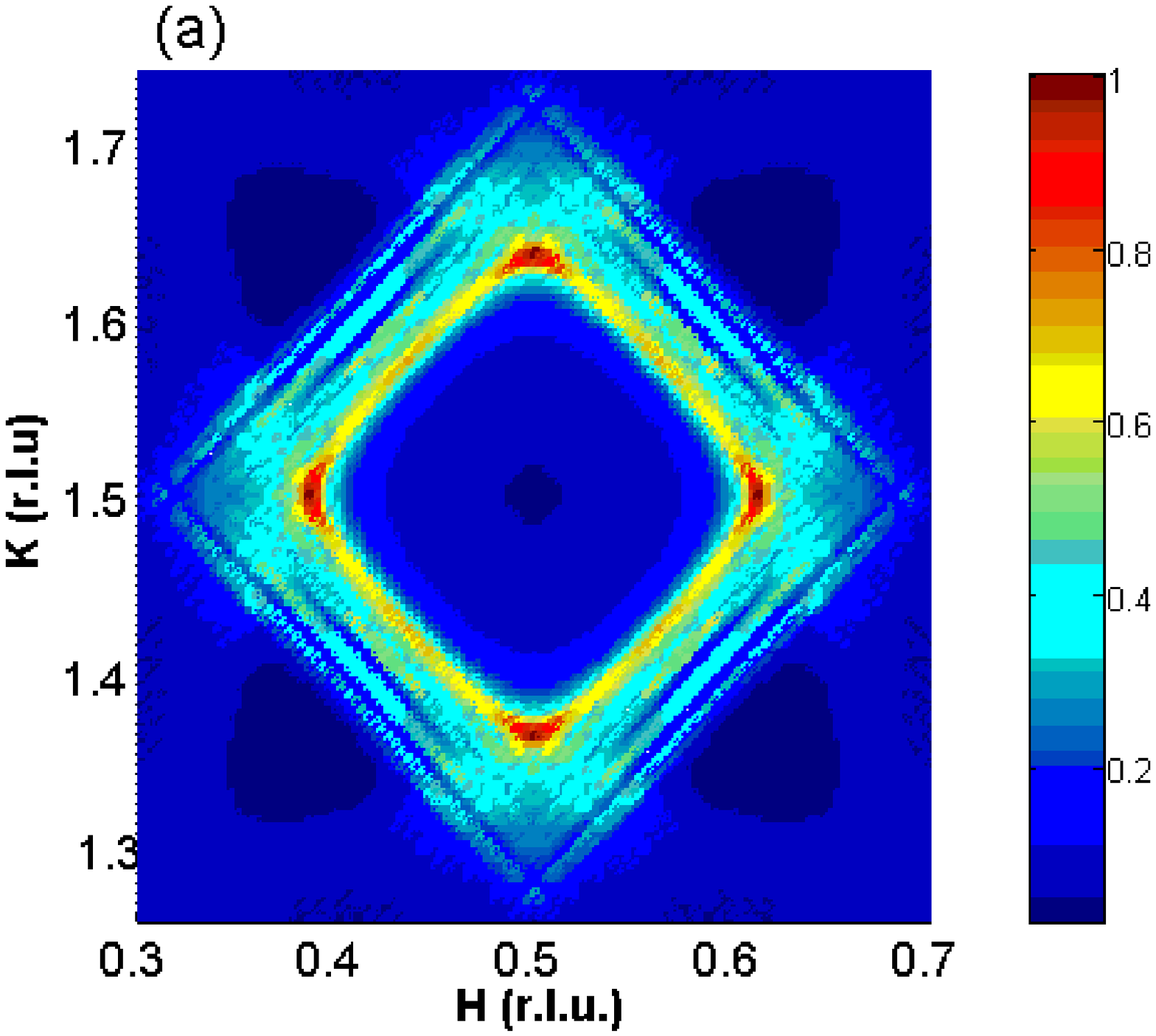,width=5cm,angle=0}}
\centerline{\psfig{file=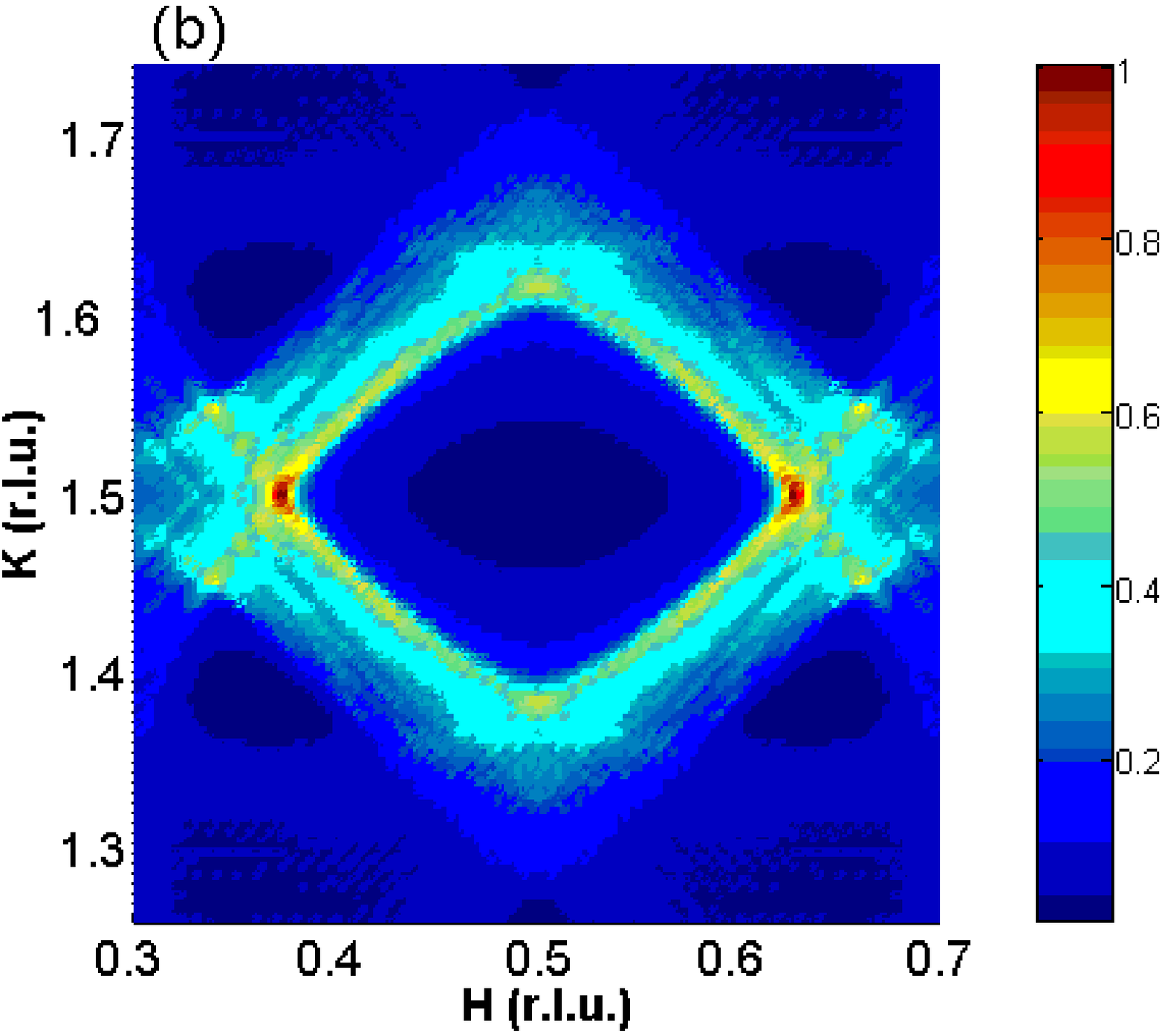,width=5cm,angle=0}}
\caption{(color) 
Calculated normalized two-dimensional intensity plot
for a constant energy of $\hbar\omega = 35$ meV, (a)
without, (b) with inclusion of orthorhombicity ($\delta_0 =0.03$). 
Note, for convenience we have chosen the axes as they are 
in the experimental work of Ref. \protect\cite{hinkov}.}
\label{twodortho}
\end{figure}

What is happening {\it below} the resonance 
threshold  
($\omega<\omega_{res}$) in fully untwinned YBCO 
for constant energy scans as a function of the 
momenta $q_x$ and $q_y$?
In Fig. \ref{twodortho} we show the calculated projected momentum
dependence of $\mbox{Im }\chi({\bf q},\omega = 35 \mbox{meV})$
without (a) and with (b) orthorhombicity. In accordance with {\it ab-initio} 
calculations, we have chosen 
$\delta_0 = 0.03$ \cite{lichti}.
In the tetragonal case one sees that the spin excitations 
form a circle around $(\pi,\pi)$ with four pronounced peaks at 
$(\pi \pm q_0, \pi)$ and $(\pi, \pi \pm q_0 )$. The origin of the peaks is 
clear: away from ${\bf Q}_{AF}$ we are connecting 
points at the Fermi surface which lie closer to the diagonal of the BZ. The 
superconducting gap tends to zero there and thus the position and the 
intensity of the resonance peak are decreasing. However, for the diagonal 
wave vectors $(\pi \pm q_0, \pi  \pm q_0)$ it happens faster than for 
the vector $(\pi \pm q_0, \pi)$ or $(\pi, \pi \pm q_0 )$. 
Therefore, effectively the latter peaks are 
'closer' to the resonance condition at ${\bf Q}_{AF} = (\pi,\pi)$; their 
intensities are higher than those for the other wave vectors. This explains 
the observed 
symmetry of the dominant spin excitations for $\omega<\omega_{res}$ 
shown in Fig.\ref{twodortho}(a).  For the
orthorhombic case the situation is changing. The ring of 
the excitations becomes distorted and, most importantly, there are only two 
well pronounced peaks. The latter is a result of strongly changed 
electronic properties, in particular, the 
topology of the Fermi surface (see inset of Fig.\ref{fig1} for comparison). 
Our {\it main result} is that, despite there are only two 
pronounced peaks, the resonant spin excitations remains basically 
two-dimensional. This is in good agreement with recent 
experiments\cite{hinkov}. Furthermore, this result based on a 
standard Fermi-liquid approach is in contrast to the stripe 
scenario of the resonance peak\cite{vojta}.

In summary, we have analyzed the in-plane magnetic anisotropy in 
high-T$_c$ superconductors with orthorhombic distortions employing a
generalized RPA-type theory and compared our results with INS data
on fully untwinned YBCO. 
We find that due to changes in the electronic structure and the 
FS topology the resonance 
peak is slightly shifted towards higher energies and that for 
$\omega < \omega_{res}$ 
the dominant spin excitations form a 2D ring-like structure 
around {\bf Q}$_{AF}$ with four pronounced peaks (tetragonal case). 
The orthorhombic distortions 
suppress two of these peaks, however, the overall 
structure of the excitations in 
Im$\chi$ remains two-dimensional which agrees well 
with recent experimental data by V. Hinkov et al.\onlinecite{hinkov}. 
Our results provides an 
alternative picture based on a conventional Fermi-liquid theory in contrast 
to the stripe scenario. 

It's a great pleasure to acknowledge  the  
enlightening discussions with V. Hinkov, B. Keimer, and Ph. Bourges. 
We also wish to thank A. Chubukov, D.K. Morr, S. Pailhes, Y. Sidis, 
D. Reznik, and K.H. Bennemann for valuable discussions. Financial support by 
INTAS (No. 01-0654) is gratefully acknowledged.

\begin{thebibliography}{10}
\itemsep 0pt
\parskip 0pt
%
\bibitem{moria} See for review 
D. Scalapino, Phys. Rep. {\bf 67}, 134520 (1995);
T. Moriya and K. Ueda, Adv. Phys. {\bf 49}, 555 (2000); 
Ar. Abanov,  A.V. Chubukov, and 
J. Schmalian, Adv. Phys. {\bf 52}, 119 (2003).
%
\bibitem{keimer} H.F. Fong, Ph. Bourges, Y. Sidis,  L.P. Regnault, 
 J. Bossy, A. Ivanov, D.L. Milius, I.A. Aksay, and B. Keimer, Phys. Rev. B 
{\bf 61}, 14773 (2000).
%
\bibitem{bourges} Ph. Bourges, L.P. Regnault, Y. Sidis, and C. Vettier, 
Phys. Rev. B {\bf 53}, 876 (1996).
%
\bibitem{emery} see for review S.A. Kivelson, Rev. Mod. Phys. {\bf 75},
1201 (2003).
%
\bibitem{vojta} M. Vojta and T. Ulbricht, cond-mat/0402377 (unpublished); 
G.S. Uhrig, K.P. Schmidt, and M. Gr\"uninger, cond-mat/0402659 
(unpublished).
%
\bibitem{seibold} G. Seibold and J. Lorenzana, cond-mat/0406589 (unpublished).
%
\bibitem{dmiekhb01} D. Manske, I. Eremin, and K.~H. Bennemann,
Phys. Rev. B {\bf 63}, 054517 (2001).
%
\bibitem{onufrieva} F. Onufrieva and P. Pfeuty, Phys. Rev. B 
{\bf 65}, 054515 (2002). 
%
\bibitem{chubuk1} Ar. Abanov, A.V. Chubukov, M. Eschrig, M.R. Norman, and 
J. Schmalian, Phys. Rev. Lett. {\bf 89}, 177002 (2002).
%
\bibitem{morr} D.K. Morr and D. Pines, Phys. Rev. Lett. {\bf 81}, 
1086 (1998).
%
\bibitem{prelovsek} I. Sega, P. Prelovsek, and J. Bonca, 
Phys. Rev. B {\bf 68}, 054524 (2003).
%
\bibitem{li} J.-X. Li, and C.-D. Gong, Phys. Rev. B {\bf 66}, 014506 (2002); 
T. Zhou, and J.-X. Li, Phys. Rev. B {\bf 69}, 224514 (2004).
%
\bibitem{hinkov} V. Hinkov, S. Pailhes, P. Bourges, Y. Sidis, A. Kulakov,
C.~T. Lin, C. Bernhard, and B. Keimer, Nature (London), in press. 
%
\bibitem{mook} H.~A. Mook, P. Dai, F. Dogan, and R.~D. Hunt,
Nature (London) {\bf 404}, 729 (2000).
%
\bibitem{norman} M.~R. Norman, Phys. Rev. B {\bf 63}, 092509 (2001); 
O. Tchernyshyov, M.R. Norman, and A.V. Chubukov, Phys. Rev. B {\bf 63}, 144507 
(2001); M. Eschrig, and M.R. Norman, Phys. Rev. B {\bf 67}, 144503 (2003).
%
\bibitem{manskefoot1} In this one-band
picture the Coulomb interaction $U$ between these quasiparticles
refers to an effective interaction within the conduction band and 
thus is smaller than the bandwidth.
%
\bibitem{gofron} K. Gofron, J.C. Campuzano, A.A. Abrikosov, M. Lindroos, 
A. Bansil, H. Ding, D. Koelling, and B. Dabrowski, Phys. Rev. Lett. 
{\bf 73}, 3302 (1994). 
%
\bibitem{schabel} M.C. Schabel, C.-H. Park, A. Matsuura, Z.-X. Shen, 
D.A. Bonn, R. Liang, and W.N. Hardy, Phys. Rev. B {\bf 57}, 6090 (1998); 
{\it ibid.} {\bf 57}, 6107 (1998).
%
\bibitem{metzner} W. Metzner, D. Rohe, and S. Andergassen,
Phys. Rev. Lett. {\bf 91}, 066402 (2003).
%
\bibitem{dmbook} D. Manske, {\it Theory of unconventional
superconductors}, Springer, Heidelberg (2004).
%
\bibitem{dmiekhb03} D. Manske, I. Eremin, and K.~H. Bennemann,
Phys. Rev. B {\bf 67}, 134520 (2003).
%
\bibitem{shen} D.~H. Lu, D.~L. Feng, N.~P. Armitage, K.~M. Shen,
A. Damascelli, C. Kim, F. Ronning, Z.-X. Shen, D.~A. Bonn, R.
Liang, W.~N. Hardy, A.~I. Rykov, and S. Tajima, Phys. Rev. Lett.
{\bf 86}, 4730 (2001).
%
\bibitem{letter01} D. Manske, I. Eremin, and K.~H. Bennemann,
Phys. Rev. Lett. {\bf 87}, 177005 (2001).
%
\bibitem{sciencebourges} P. Bourges, Y. Sidis, H.~F. Fong, L.~P. Regnault,
J. Bossy, A. Ivanov, and B. Keimer, Science {\bf 288}, 1234
(2000).
%
\bibitem{lichti} O.K. Andersen, A.I. Liechtenstein, O. Jepsen, 
and F. Paulsen, J. Phys. Chem. Solids {\bf 56}, 1573 (1995).
%
\end{thebibliography}
\end{document}